\documentclass[journal]{IEEEtran}
\usepackage{cite}
\usepackage{upgreek}
\usepackage{algorithm}
\usepackage{caption}
\usepackage{etoolbox}
\usepackage{xcolor}
\usepackage{tabularx}

\usepackage{subcaption}
\usepackage{etoolbox}
\usepackage{amsmath,amssymb,amsfonts}
\usepackage{algorithmic}
\usepackage{graphicx}
\usepackage{textcomp}
\usepackage{wrapfig}

\usepackage{textcomp}
\usepackage{wrapfig}

\makeatletter
\def\ps@IEEEtitlepagestyle{%
  \def\@oddfoot{\mycopyrightnotice}%
  \def\@evenfoot{}%
}
\def\mycopyrightnotice{%
  {\footnotesize  \textcolor{red} {This paper has been accepted for publication by the IEEE Internet of Things Journal. The copyright is with the IEEE.} \hfill}
  \gdef\mycopyrightnotice{}
}

\begin{document}

	\title{Hash-MAC-DSDV: Mutual Authentication for Intelligent IoT-Based Cyber-Physical Systems}

	\author{Muhammad~Adil~(\IEEEmembership{Student,~Member,~IEEE}), Mian~Ahmad~Jan$^{*}$ (\IEEEmembership{Senior~Member,~IEEE}),
		Spyridon~Mastorakis~(\IEEEmembership{Member,~IEEE}), Houbing~Song~(\IEEEmembership{Senior~Member,~IEEE}),
		Muhammad~Mohsin~Jadoon \IEEEmembership{(Member, IEEE)},
		Safia~Abbas, Ahmed~Farouk~(\IEEEmembership{Member,~IEEE})
		
		\thanks{Muhammad~Adil is with the Department of Computer Science, Virtual University of Pakistan, 54–Lawrence Road, Lahore 54000, Pakistan (email:ms170401318@vu.edu.pk).}
		\thanks{Mian~Ahmad~Jan is with the Department of Computer Science, Abdul Wali Khan University Mardan, Pakistan (mail:mianjan@awkum.edu.pk).}
		\thanks{Spyridon Mastorakis is with the Computer Science Department, University of Nebraska at Omaha, Omaha, USA (email:smastorakis@unomaha.edu).}
		\thanks{Houbing~Song is with the Department of Electrical Engineering and Computer Science, Embry-Riddle Aeronautical University, USA (Email:h.song@ieee.org).}

		\thanks{Muhammad Mohsin Jadoon is with the Department of Radiology	and imaging, Yale University, New Haven, CT, USA and Department of Electrical Engineering, International Islamic University Islamabad (Email address:muhammedmohsin.khan@yale.edu/mohsin.khan@iiu.edu.pk)}
		\thanks{Safia~Abbas is with the Computer Science Department, Princess NourahBint Abdulrahman University, KSA (email:samahmoud@pnu.edu.sa).}
		\thanks{Ahmed~Farouk is with the Computer Science and Physics Department, Wilfrid Laurier University (email:Afarouk@wlu.ca).}
		
		\thanks{Computer Science and Physics Department, Wilfrid Laurier University (Email address:Afarouk@wlu.ca)}
		
		\thanks{Corresponding author:Mian Ahmad Jan}}

	\markboth{IEEE journal of Internet of Things,}%
	{Shell \MakeLowercase{\textit{Adil et al.}}: Bare Demo of IEEEtran.cls for Journals}
	
	\maketitle
	
	\begin{abstract}
		Cyber-Physical Systems (CPS) connected in the form of  Internet of Things (IoT) are vulnerable to various security threats, due to the infrastructure-less deployment of IoT devices.  Device-to-Device (D2D) authentication of these networks ensures the integrity, authenticity, and confidentiality of information in the deployed area.  The literature suggests different approaches to address security issues in CPS technologies. However, they are mostly based on centralized techniques or specific system deployments with higher cost of computation and communication. It is therefore necessary to develop an effective scheme that can resolve the security problems in CPS technologies of IoT devices. In this paper, a lightweight Hash-MAC-DSDV (Hash Media Access Control Destination Sequence Distance Vector) routing scheme is proposed to resolve authentication issues in CPS technologies, connected in the form of IoT networks. For this purpose, a CPS of IoT devices (multi-WSNs) is developed from the local-chain and public chain, respectively. The proposed scheme ensures D2D authentication by the Hash-MAC-DSDV mutual scheme, where the MAC addresses of individual devices are registered in the first phase and advertised in the network in the second phase. The proposed scheme allows legitimate devices to modify their routing table and unicast the one-way hash authentication mechanism to transfer their captured data from source towards the destination. 
		Our evaluation results demonstrate that Hash-MAC-DSDV outweighs the existing schemes in terms of attack detection, energy consumption and communication metrics. 
	\end{abstract}
	
	
	\IEEEpeerreviewmaketitle

	\section{Introduction}
	
	\IEEEPARstart{I}{nternet}  of Things (IoT) is the latest and emerging trend in the current era of Information and Communications Technology (ICT). IoT has numerous applications in the real world that include disaster management, military surveillance, healthcare, smart farming, and industrial automation among others [1-3].
	The network architecture of IoT is mainly based on distributed and centralized communication infrastructures \cite{gubbi2013internet, khan2020secured, jan2020security, mastorakis2020icedge}.  In the distributed infrastructure, the clients directly extract data from the deployed sensor devices, while in the centralized  infrastructure, the sensor devices in the deployed area process the data collected via a concerned base station.  
	
	For the IoT  connectivity, multiple Wireless Sensor Networks (WSNs) collaborate to deliver services to the end users. It is therefore essential for these networks to manage the identity of sensor devices in a secured environment \cite{khan2018iot, zhong2020automated, jan2020lightweight, shannigrahi2020next}. In most cases, the deployment of WSN and IoT infrastructure is challenging, so Device-to-Device (D2D) authentication is a viable option in these  scenarios. In the existing literature, the D2D authentication schemes used for IoT  are mostly used in a centralized fashion  \cite{mahmod2020robust, ferrag2017privacy}. However, in a centralized D2D authentication, legitimate sensor devices rely on third parties, e.g. authentication servers, for their verification and network participation. This increases the likelihood of failures since all participating devices depend on a single point for their authentication. 
	The distributed or decentralized authentication resolves this issue in these networks. Authentication of IoT networks connected in the form of multi-WSNs is an emerging decentralized approach used to address the issues associated with a centralized authentication  [8]-[11].

	The interconnection of multi-WSNs brings numerous challenges in terms of network security, architecture, lifetime, and communication metrics, due to its continuous emergence~\cite{ullah2020icn, mastorakis2020dapes, rehman2020ccic, li2019distributed}. Data confidentiality and integrity are the crucial aspects of IoT networks because they ensure the legitimacy of a network. Therefore, D2D authentication of IoT networks interconnecting multi-WSNs is mainly focused on the topological structure and routing protocols to create a secured communication infrastructure.  The existing  techniques use  the  peer-to-peer  authentication  through   nodes,  servers  or base  stations, which is mostly based on centralized communication. In addition, the centralized communication of these sensitive networks is not reliable because failure of a centralized device may disrupt the operation of  interconnected sites and may create network overhead in terms of shifting complete load to the neighboring servers or controllers, which degrades the network performance. To resolve the authentication issue in these networks, we have proposed a decentralized approach in this research, which is effective in terms of various authentication modes. The main contributions of this research are as follow.
	
	
	\subsection{Research contributions}
	
	In this paper, we propose a Hash-MAC based mutual authentication scheme for Cyber-Physical Systems (CPS) with embedded sensor devices connected in the form of multi-WSNs topology of IoT. Unlike the existing studies, we focus on decentralized authentication, where a D2D approach is adopted to verify the authenticity of participating devices. Moreover, the proposed model uses a hash function with MAC addresses to create a secured authentication key for the deployed IoT devices. We have developed the IoT network infrastructure by interconnected multi-WSNs, which are further built into various components like base station (BS), cluster head (CH) and ordinary sensor devices. The legitimate devices of the network register their MAC addresses in a local chain with a concerned CH node followed by BS in the public chain. In the next step, the BS broadcasts the MAC addresses of registered devices in the public chain by adding a hash function. Likewise, BS(s) connected in the public chain receives this information and forwards it to the local chain through CH nodes followed by legitimate devices. Consequently, the embedded devices of CPS in the local chain update their routing table and follows this information for unicast communication or one-way authentication in the network.
	 \newline The major contributions of our scheme are as follows: 
		\begin{enumerate}
			\item	To reduce the computation and communication costs, the CH nodes and BS(s) have been provided with sufficient power and control to manage the MAC address registration of legitimate devices, hash functioning and route advertisement for creating a congestion-free communication environment with a better lifespan of legitimate devices. 
			\item	 The authenticity of participating devices is ensured by continuously advertising their authentication information in the local and public chain. In addition, the legitimate devices in the local chain follow the advertised information with a one-way Hash MAC authentication process to ensure the legitimacy of requesting devices by matching Hash MAC with their MAC address table.
			\item	The D2D authentication is achieved by utilizing minimal network resources in our proposed model, because most of the computation is performed by BS and CH nodes, which improve the result statistics in terms of comparative metrics over the existing schemes. In addition, the network overhead is minimized up to a remarkable level in terms of throughput, packet lost ratio, and network lifetime, because the participating device follows only the advertised information of the concerned CH node and BS, respectively, to send their data from source to destination.
\end{enumerate}
Beside that, we want to acknowledge that our proposed model presents a secured communication environment with robust MAC address registration to ensure their legitimacy. Thus, the proposed HASH MAC scheme applies to any IoT application that has legitimate device authentication and secure communication requirements e.g. industrial automation, military, internet of vehicles (IoV), agriculture, smart homes etc. \\		
The rest of the paper is organized as follows. Section 2 presents the related work followed by the proposed model of authenticating the IoT devices in CPS in Section 3. Section 4 presents the formal security analysis followed by  experimental results in Section 5. Finally, section 6 concludes our paper.

		\section{Literature Review}
		Wireless Sensor Networks (WSNs) and Internet of Things (IoT) are susceptible to various security threats, due to the deployed environment and their dynamic communication behavior.  Therefore, security of these networks is considered as primary concerns for research community that needs to be implemented at the deployment stage \cite{yu2019wireless}\cite{ adil2020anonymous}.  In order to combat security threats faced by these networks, the limited resources and critical applications of IoT require the research community to devise new techniques or modify existing techniques.  Therefore, the literature suggests different schemes for countering various type of attacks faced by these networks.

		Khalid et al. \cite{khalid2020decentralized} proposed a decentralized authentication scheme for IoT-based communication infrastructure of multi-WSNs, where they used fog computing as a public  authentication process to verify the legitimacy of participating sensor devices. Tonyali et al. \cite{tonyali2018privacy} proposed a privacy preserved protocol for wireless mesh networks of Smart Grid to collect data in a secured communication environment. Moreover, they used Fully Homomorphic Encryption (FHE) and Secure multiparty Computation (SMPC) techniques in their model to aggregate data in a secured communication environment.  A comprehensive survey of secure services composition and data aggregation for WSNs was carried out by Aloqaily et al. \cite{aloqaily2019data}. Baker et al. \cite{baker2020secure} proposed the toolbox authentication scheme for blockchain wireless sensor networks. In the proposed model, they used a key feature for the authenticity and identity of communicating sensor devices. Moreover, they used the signature and cryptographic approaches in their cloud-based infrastructure of blockchain wireless networks. Rathee et al. \cite{rathee2019blockchain} proposed a blockchain authentication scheme for autonomous vehicular communication infrastructure.

		The challenges associated with the security of IoT network communication infrastructure was expansively overviewed by Tariq et al. \cite{tariq2019security} in their survey article. Moreover, the authors comprehensively discussed the well-established approaches to counter these threats in an operational network. However, the survey was adopted specifically against fog computing IoT network communication environment.   The two-factor lightweight privacy-preserving authentication scheme for IoT devices connected in the form of multi-WSNs was proposed by Gope et al. \cite{gope2018lightweight}. The authors used the physical functionality of IoT devices as an authentication factor to verify the legitimacy of participating devices. However, later on, their scheme was flawed, because every time the assessment of the physical properties of IoT devices is not possible. Feng et al. \cite{feng2018aaot} was proposed the lightweight attestation and authentication scheme for the IoT network. They used the memory and clone-able functions of sensor devices with the help of software-define infrastructure to verify the authenticity of participating devices in the network. The CreditCoin based authentication scheme for blockchain vehicular wireless sensor networks was proposed by Li et al. \cite{li2018creditcoin}. The limitation of this scheme was its specific system implementation with a complex model.

		Cui et al. \cite{cui2018detection} used a deep learning technique to identify malware variants in the deployed WSNs. In the proposed scheme, the authors used malicious codes in the conversion of grayscale images to verify the legitimacy of their scheme. The scheme was effective to identify malware variants in the network, but it was limited to homogeneous networks, which minimizes its use in the real deployment. Aitzhan et al. \cite{aitzhan2016security} proposed a third party decentralized authentication scheme for smart grid energy system. Moreover, the authors used blockchain technology infrastructure in their model with multi-signature anonymous encryption message streams to ensure the security of deployed IoT.  A detailed survey on blockchain technology in coordination of a centralized approach to overview the security performance of deployed WSNs infrastructure is carried out by Salman et al. \cite{salman2018security}.\\
		Cui et al. \cite{cai2019ensemble} was proposed the ensemble bat algorithm (BA) approach for large scale optimization problems by integrating ideas. Edge Chain named blockchain communication environment for IoT networks was proposed by Pan et al. \cite{pan2018edgechain}.  The basic idea of the proposed model was to integrate the blockchain of WSNs and link them through edge cloud for secure information exchange.  The three-tier security architecture for IoT networks was suggested by Bao et al. \cite{bao2018iotchain}. The authors used the blockchain layer, authentication layer, and application layer in combination to resolve the security issues of IoT networks.  Won et al. \cite{won2017certificateless} was proposed the Certificateless Signcryption Tag Key Encapsulation Mechanism (eCLSC-TKEM) to resolve the security issue of city-based drone communication infrastructure. The proposed model was capable to verify the authenticity of participating devices in terms of relationships such as one-to-one authentication, one to many, and many to one.
		
		\subsection{Limitation of Existing Schemes.}
		CPS is extremely vulnerable to a variety of internal and external attacks, due to their unrestricted area deployment and complex communication activities. As a consequence, efficient use of CPS networks enriches their productivity and adaptability. CPS is made up of hundreds of thousands of sensor nodes that are exposed to numerous forms of attacks, i.e. jamming, black hole attacks, Sybil attacks, server-side attacks, and so on. Various methodologies had been demonstrated in the recent past to mitigate these types of attacks and resolve the authentication problem in CPS, but the majority of them are specific to the communication environment, system deployment, or software relevancy. \\
			The following are some of the major disadvantages correlated with the current literature:
			\begin{enumerate}
				\item The majority of the authentication approaches discussed in the literature are difficult to execute in a real deployment, due to their complexity.
				\item Some of the discussed techniques are successful for unique network attacks, which limits their use in real-world applications because CPS is susceptible to numerous network attacks.
				\item Some present literature employs a convoluted authentication mechanism, which adds to network overhead and reduces network capability in terms of computation and communication costs.
			\end{enumerate}
			\section{Proposed Methodology: Hash-MAC-DSDV Scheme}\label{Overview}
			
			
			\subsection{Assumptions and System Model}
			
			Let us assume an IoT network of sensor devices, CH nodes, and BSs. The sensor devices have limited resources in terms of storage, transmission, processing, and onboard power. Such resources require efficient utilization for better results and prolonged network lifetime. To this end, in our proposed model, most computation is performed by CH nodes and BS(s) to maximize the lifetime of sensor devices. 
			The sensor devices are components of CPS connected in the form of IoT networks. Moreover, these devices are deployed at a designated location to collect and process information in the network according to their assigned task. 
			The CH nodes process the gathered information of sensor devices in the local chain and transmit it to the BS(s) for further processing. Each CH is a special device with higher processing and memory capabilities, as well as onboard power. Therefore, they process the data in an effective way in the local chain as well as in the public chain associated with the BS. The BS connects multi-WSNs of CPS to form a heterogeneous network. Moreover, the BS works as a point of interest for sub-networks, because it manages the sub-networks in terms of security and data processing. 
			
			\begin{figure*} [h!]
				\centering
				\includegraphics[width=0.6\linewidth, height = 13. cm]{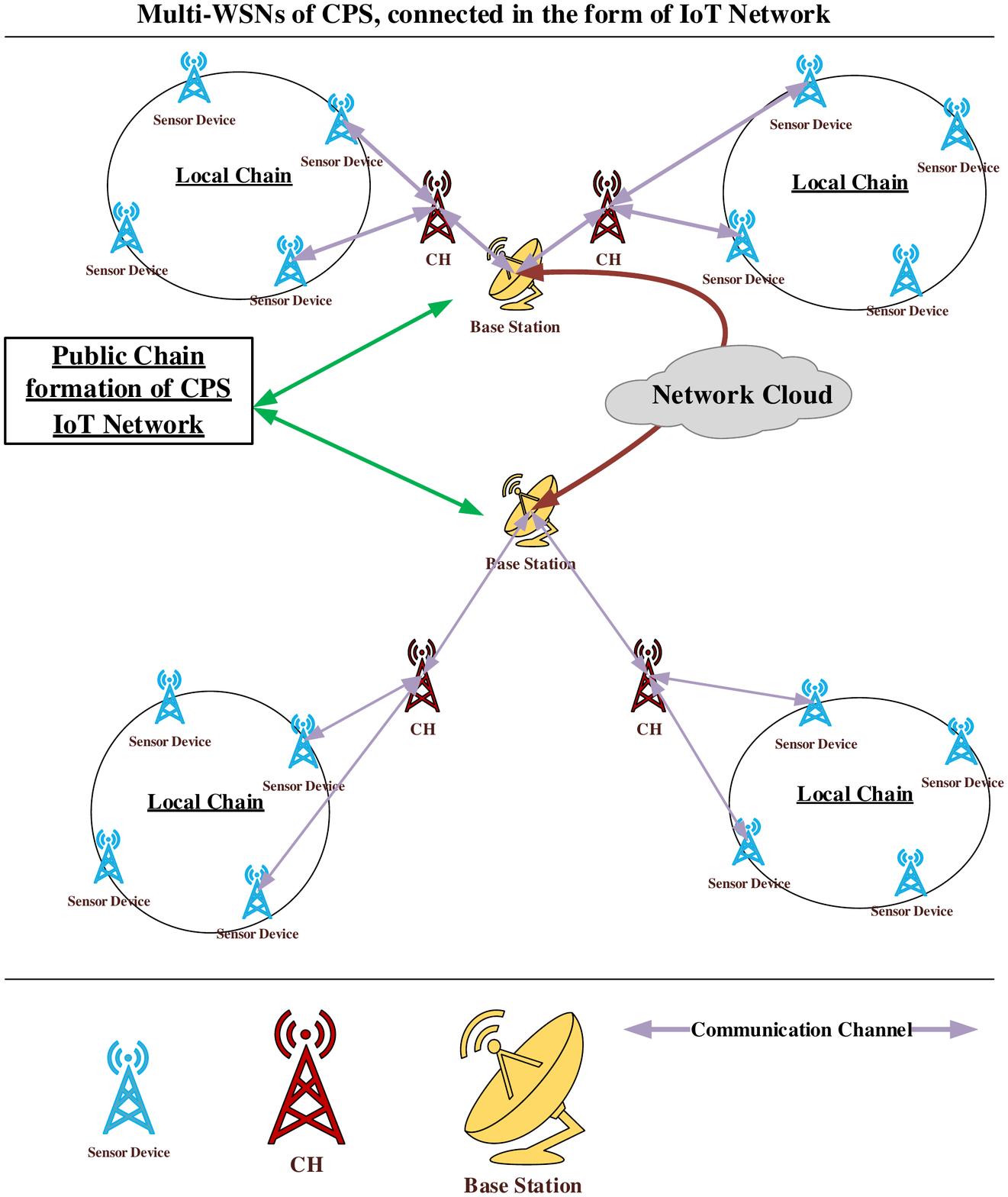}
				\vspace{-1cm}
				\caption{Architecture overview of local and Public chain of our scheme}
				\label{fig : 1}
			\end{figure*}
			
			Figure \ref{fig : 1}  represents the architectural diagram of our proposed approach. The connectivity between legitimate devices and cluster heads is shown by light blue lines whereas, the green lines  show intra-connectivity among the BSs, while the orange lines represent the connectivity of BSs and network cloud. The black lines in the figure show the D2D communication among different clusters.  
			The sensor devices in each cluster gather data according to their designated task and process it via the concerned CH in the network. Similarly, each CH node uses an associated BS to transmit the gathered information to remote destinations within the network. Moreover, as shown in the figure, the legitimate devices of one cluster can communicate with another cluster's device, if they verify the security parameters of the proposed model by matching the MAC addresses. The multi-WSNs of CPS, as shown in the figure,  establish a heterogeneous IoT network. 
			
			\subsection{Authentication Mechanism}
			
			The authentication mechanism adopted in the proposed model includes the following phases: initialization, registration, and authentication.
			
			\subsubsection{Initialization and Registration Phases}
			
			The initialization phase is mainly concerned with the BSs as they initiate the Hash-MAC-DSDV mutual authentication scheme on sub-networks using the local and public chain connectivity. All the participating sub-network devices first register their MAC addresses with a BS in an offline phase within the public chain va a concerned CH (local chain). The BS adds MD5 Hash function with the registered MAC address and broadcasts it within the public chain following the Hash-MAC-DSDV routing scheme. The CH connected with the BS passes on this information to the local chain, where the legitimate devices update their routing table according to the advertised information. Similarly, the connected devices in the network follow their routing table information to transmit data from source to destination via concerned CH and BS. However, before transmission, a one-way-authentication process is carried out to verify the legitimacy of each requesting device by matching its MAC address in the device MAC table, CH, or BS, respectively.
			
			\textbf{Theorem-1:} An ordinary device $D_i$ generates a  MAC address registration request with concerned $S_j$ through CH, where $D_j$ is the specified BS. $D_i $ communicate in the network, if $D_i$ MAC address $\in$ $(D_j)$ public chain.
			\newline\textbf{Proof:} Let us assume that an attacker device $A_k$ generates a registration request directly with BS by sending its MAC address.  $D_j$ checks the MAC address of requesting device by triggering a lookup with connected CH or local chain network. After, a through check, the MAC-address of $A_k$  did not match with any local chain of  $S_j$. Hence, the MAC address of $A_k$   cannot be registered by concerned $D_j$ in the public chain. Moreover, the registration request of $A_k$ is denied by $D_j$.
			\newline Conversely, if the legitimate device $D_i$ generated a registration request with the concerned $D_j$ through CH or local chain, then its MAC address will be verified successfully in terms of $D_i \in member(D_j,i)$ , where the $i^{th}$ term in $D_{j,i}$ denotes the total number of registered MAC addresses of legitimate devices  in $D_{j}$, which is further classified as  i=(1,2,3,4,.......n-1), in our proposed multi-WSNs network.  Hence,  $D_j$  registers only those devices that approached through local chain or CH in the initial phase of registration (offline phase).
			\begin{algorithm}[htbp]
				\caption{Registration of ordinary devices $D_i$ with  $D_j$ via the specified CH}
				\label{alg : 1}
				\begin{algorithmic}[1]
					\REQUIRE Registration of legitimate $D_{i}$ with concerned $D_{j}$.   
					\ENSURE Registration of legitimate $D_{i}$ in the local and public chains.  
					\STATE $D_i$ generate registration RREQ with $D_j$  \\
					\STATE $D_{i}$  forward registration RREQ  $\longleftarrow$  through CH \\	
					\STATE 	$D_{j} $ $\longleftarrow$ Receives $D_{i} $ RREQ through CH, where, i=1,2,3......n-1
					\STATE \quad \textbf{for} (i=0; i = n-1; i++)\\
					
					\STATE \qquad $D_{j}$ check local chain (CH)  $\longleftarrow$ of $D_{i}$ \\ 
					\STATE \quad \qquad \textbf{if} \\
					\STATE \quad \qquad $D_{i}$ RREQ $\in$ local chain of $D_{j}$    \\
					
					\STATE \quad \qquad \textbf{then,}
					\STATE \quad \qquad $D_{j}$ registers $D_{i}$ MAC address in their MAC table
					\STATE \quad \qquad \textbf{Else}
					\STATE \quad \qquad $D_{j}$ Denies $D_{i}$ registration request
					\STATE \qquad \textbf{end if}  
					
					\STATE \qquad  $D_{j}$ broadcast the registered MAC address in the public chain
					\STATE \quad CH $\longleftarrow$ Share information of public chain in local chain
					\STATE \quad ordinary devices update their routing table according to $D_{j}$  
					
					\RETURN: List of registered $D_{i}$ devices MAC addresses 
					
				\end{algorithmic}
			\end{algorithm}

			Algorithm \ref{alg : 1} describes the registration phase of the proposed model. Initially, $D_{i}$  devices generate RREQ messages to register their MAC addresses with $D_{j}$ that checks the local chain of requesting device. If the local chain (CH) $\in$ $D_j$, then, the $D_j$ registers the MAC addresses of requesting devices, else they deny the registration request of these  devices. Likewise, the $D_j$ advertises the information of registered devices in public chain through connected $D_j$ and local chain through CH nodes for other legitimate connected devices. All connected devices in the local chain and public chain update their routing table for communication in the network. Finally,  $D_{j}$ represents the list of registered MAC Addresses. Please note that only registered devices can process their request through concerned CH in the public chain.

			
			\begin{figure*} [h!]
				\centering
				\includegraphics[width=0.6\linewidth,  height = 6.7cm ]{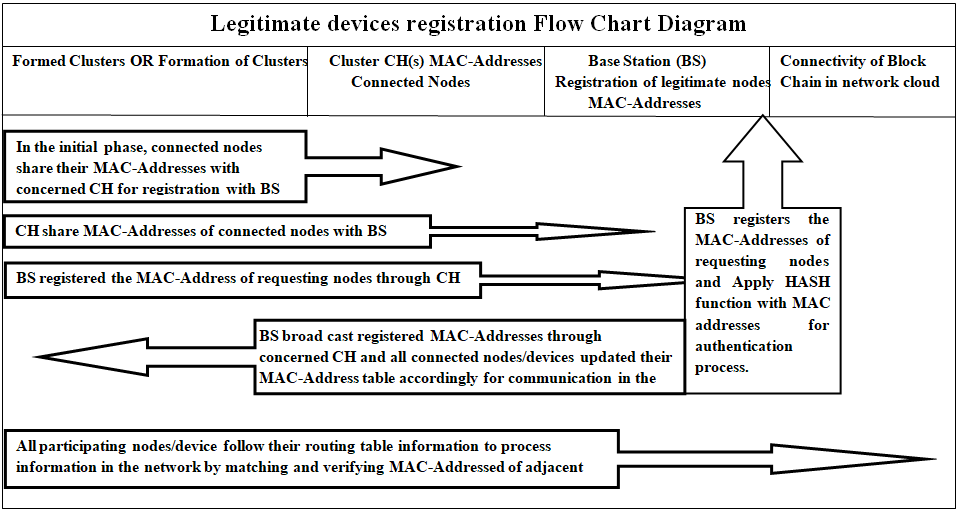}
				\caption{Legitimate devices registration step by step process in the proposed model}
				\label{fig : 2}
			\end{figure*}
			
			The flow chart of Figure~\ref{fig : 2} depicts the registration process of our proposed Hash-MAC-DSDV algorithm. The participating devices $D_{i}$ share their MAC addresses with $D_{j}$ through a CH, which we call local chain in the proposed model. 
			However, once $D_{j}$ receives a registration request from a CH for an ordinary device $D_{i}$, the $D_{j}$ registers the MAC address of requesting $D_{i}$ and uses the MD5 hash algorithm to broadcast the MAC address of $D_{i}$ in the network via other $D_j$ interconnected small WSNs, consequently, the $D_{j}$ further broadcast this information to connected CH nodes in the local chains. 
			When the participating devices in the local chain receive the CH broadcast message, they update their routing table. Similarly, these devices follow their routing information to transmit messages in the network and make authentication possible between D2D in the local chains, device to CH, and CH to $D_{j}$ by means of the proposed scheme. 
			To elaborate further, the devices connected in the local chain follow their routing table information to forward or receive communication requests. Moreover, these information contains the adjacent devices hop-count, distance, and MAC address etc. which ensures the effectiveness of proposed scheme. 
			Likewise, the $D_{j}$ maintains a record of the local chains in terms of clusters (CH) with their $D_{i}$ devices MAC address information and the CH maintains a record of legitimate devices. Consequently, the connected $D_{i}$ of the network follows the rule of Hash-MAC-DSDV protocol for communication in the network.
			
			\subsubsection{Authentication Phase}
			
			Let us assume that a legitimate device $D_{i}$, where $D_{i}$ $\in$ local chain of the CH, initiates an authentication request with $D_{j}$ to process the collected data in the public chain. $D_{i}$  processes its data through the local chain using the CH, which checks and matches the requesting device ID in terms of MAC addresses in their MAC table to verify its authenticity in the local chain. If MAC address of ($D_{i}$)  $\in$ CH list, then the CH processes the $D_{i}$'s request for further processing in the network. Once $D_{j}$ receives $D_{i}$'s request for communication in the network, $D_{j}$ matches $D_{i}$'s MAC address in its routing table by following the public chain mechanism for the specified CH (local chain). $D_{j}$  matches the requesting $D_{i}$'s MAC address in its routing table, if $D_{i}$'s MAC address $\in$ $D_{j}$ MAC address list for a specified local chain, $D_{j}$ allows $D_{i}$ to communication through the public chain. Hence, $D_{j}$  verifies the authenticity of requesting $D_{i}$ devices and processes their information in the network. 
			
			\begin{figure*} [h!]
				\centering
				\includegraphics[width=0.6\linewidth, height = 6.8cm ]{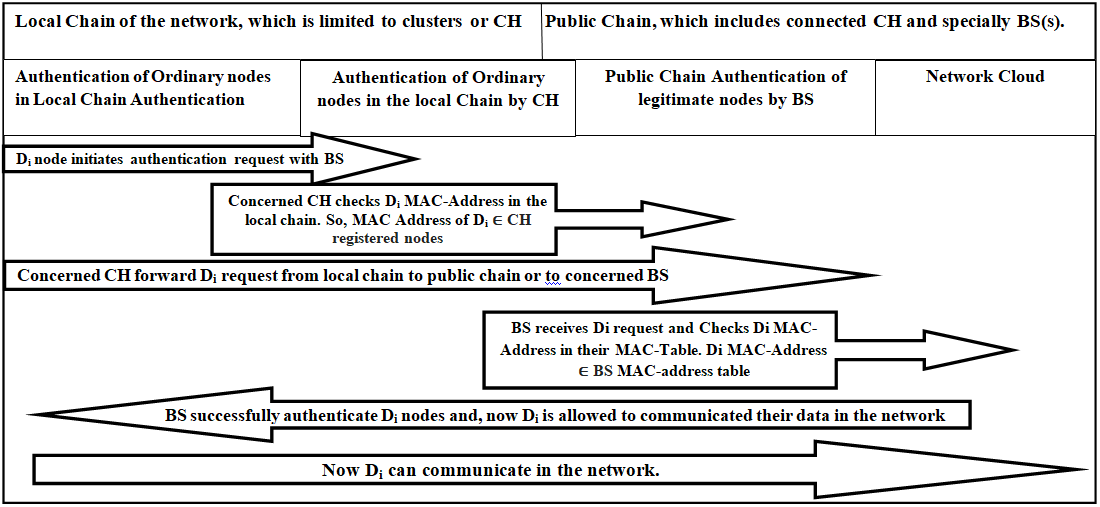}
				\caption{Authentication Flow chart diagram of Hash-MAC-DSDV protocol}
				\label{fig : 3}
			\end{figure*}
			
			Figure~\ref{fig : 3} illustrates the step by step authentication  for a device registered in the network. The proposed model works on the basis of  chains:  local and  public. Similarly, a number of small WSNs of CPS are interconnected in the network to form a heterogeneous IoT network. $D_{i}$ initiates a message exchange request in the network. $D_{i}$ needs an authentication in the local chain to move forward in the public chain. Therefore, $D_{i}$'s message contains information such as device ID and MAC address. This information is checked by the CH in the local chain to verify its authenticity. The $D_{i}$'s message is verified by the CH via matching its MAC address in the local chain. If $D_{i}$'s MAC address $\in$ CH's MAC address list, then the CH successfully authenticates $D_{i}$ and processes $D_{i}$'s request in the public chain. Likewise, $D_{i}$'s message request goes to the public chain, where $D_{j}$ checks and matches the MAC address of $D_{i}$ in its MAC address table. If $D_{i}$'s MAC address $\in$  $D_{j}$ MAC address table, the authentication of $D_{i}$ is completed successfully and it is allowed to communicate in the network.  
			
			\subsection{Device to Device Authentication in Local Chain}
			
			The proposed model is effective in D2D authentication because all the MAC addresses of legitimate $D_{i}$ are broadcast by the respective $D_{j}$ in the public chain. The connected CH nodes pass this information to the local chain and the participating devices in the local chain update their routing table according to the advertised information of $D_{j}$. The devices follow their routing information to communicate with neighboring devices or process their collected data through these devices and CH nodes, if they are at the next hop.
			
			Let us assume that $D_{i}$ devices generate an authentication or message exchange request with another $D_{i}$ $\in$ $D_{n-1}$, where the $i^{th}$ term represents the number of legitimate devices in the local chain such that, i=(1,2,3,4,......,n-1) devices. The receiving device checks the MAC address of $D_{i}$ in its MAC address table. If $D_{i}$'s MAC address $\in$ $D_{n-1}$ MAC address list, the receiving device will process $D_{i}$ request in the local chain. Otherwise, $D_{n-1}$ device denies $D_{i}$'s authentication request and broadcasts an alarm message in the network to acknowledge the existence of a malicious device \textit{A$_k$})in the network. 
			
			\subsection{Authentication of Attacker Devices}
			
			The authentication of an attacker device in an operational network ensures the reliability and performance of a protocol or security scheme. Therefore, the proposed Hash-MAC-DSDV scheme is effective in terms of identifying malicious devices in the  network.
			
			\subsubsection{Authentication of attackers in the local chain in terms of D2D} \hfil
			
			Let us assume that an attacker  $A_{k}$ initiates an authentication request. $A_{k}$'s request is received by a legitimate device $D_{i}$ in the cluster (local chain), where the $i^{th}$ term i= (1,2,3,4,......,n-1), represents the legitimate devices in the cluster. Once $D_{i}$ receives $A_{k}$'s message request, it matches the MAC address of $A_{i}$ in their MAC address table. Likewise, if $A_{k}$'s 
			MAC address $\notin$ $D_{i}$'s MAC address, $D_{i}$ denies $A_{k}$'s authentication request in the local chain and avoid attacks in the local chain.
			
			\subsubsection{Authentication of attackers in the local chain by CH} \hfil
			
			It is a major concern when  $A_{k}$ initiates a message request with a CH node in the local chain to compromise  its security. However, in  case of direct communication with CH nodes, when a CH receives $A_{k}$'s message request, it checks the MAC address of $A_{k}$ in its MAC address list. If $A_{k}$'s MAC address $\in$ CH MAC address list, the CH allows $A_{k}$ to communication in the network. Otherwise, CH denies $A_{k}$'s request and advertises an alarm message to notify others about the existence of an attacker in the network. 
			
			\subsubsection{Authentication of attackers in the local chain by $D_{j}$} \hfil
			
			There is a possibility that $A_{k}$ will compromise the security of the public chain by communicating directly with a $D_{j}$. 
			The proposed model is effective to combat a direct connectivity request from $A_{k}$  to $D_{j}$ through accurate identification. Once the $A_{k}$  generates an authentication request to  $D_{j}$ directly in public chain, $D_{j}$ matches $A_{k}$'s MAC address in their MAC address list. If $A_{k}$'s MAC address $\notin$ $D_{j}$ MAC address list, $A_{k}$ is accurately identified and detected by $D_{j}$. At this stage,  $D_{j}$ broadcasts an alarm message in the local chain through CH nodes to acknowledge the existence of a malicious device in the network's public and local chains.
			
			\section{Evaluation}
			
			The feasibility of  Hash-MAC-DSDV scheme was evaluated  using OMNeT++. The parameters used to implement our scheme are shown in Table \ref{tb1}. Although, we have evaluated our scheme in a simulation environment, the results of OMNeT++ are an approximation of the real environment in terms of operation. Throughout our simulation study, we tested numerous criteria for the proposed scheme to check its feasibility for the real-world implementation. Moreover, the results are validated through  formal safety analysis in the context of different threats to the network. 
			We  computed the computation and communication cost, packet losses, and latency, respectively. In addition, we compared the energy consumption of Hash-MAC-DSDV to the legacy DSDV routing protocol. 
			A description of our findings is presented in the following subsections.

			\begin{table}[h!]
				\caption{Parameters used for Hash-MAC-DSDV setup.}
				\begin{center}
					\begin{tabular}{  c   |  c  }	
						\hline \textbf{ Parameter Description } & \textbf{Value of the parameters} \\
						Sensor Devices  &   300, 600, 1000, 1500, 2000  \\
						Number of Base Stations  & 3 , 6, 10. 15, 20 \\
						Routing Protocol & Hash-MAC-DSDV \\
						Number of Cluster Heads  & 15, 30, 50, 60, 90 \\
						Initial Energy of devices ($ E_{i} $) & 60,000 mAh \\
						Simulation Tool  & OMNeT ++   \\
						
						Energy Consumption during Normal state  & 1.03 mW  \\ 
						Energy Consumption during transmission   & 70.1 mW \\
						Energy Consumption during Sleep mode  & 0.50 $ \mu $ W  \\
						
						Transmission interval of devices  & 14 $\mu$Sec \\
						
						Energy Consumption during reception   & 44.6 mW \\
						
						Residual Energy  of a device ($ E_{r}$) & 	$ E_{i} $ - $ E_{c} $ \\
						Network Traffic type & UDP \\
						Packet Size & 128 Bytes \\ 
						Communication Pattern & broadcast/ unicast \\
						
						\hline
					\end{tabular}
				\end{center}
				\label{tb1}
			\end{table}
			
			\subsection{Formal Security Analysis of Hash-MAC-DSDV}
			
			
			In this section, we first evaluate different threats and analyze our scheme by comparing against the existing ones. A brief overview of formal security analysis is shown in Table~\ref{tb2}.
			
			\subsubsection{Eavesdropping Attack}
			
			In an eavesdropping attack, an adversary $A_{k}$  steals sensitive data transmitted through an insecured communication channel. Assume that a legitimate device $D_{i}$ $\in$ $D_{n-1}$ transmits data through the local chain.  $A_{k}$ tries to capture this data over the communication channel and access the information. In our model,  $A_{k}$  needs $2^{128}$ iterations to access a message digest and $2^{512}$ iterations to access the block of messages, which is virtually impossible for sensor devices due to their limited computing and memory resources, as well as onboard power. Therefore, the Hash-MAC-DSDV scheme efficiently safeguard against the eavesdropping attack.
			
			\subsubsection{Sensing Device Impersonation Attack} 
			
			In this form of attack,  $A_{k}$ impersonates as a valid device on the network. However, our Hash-MAC-ASDV scheme is effective against this attack because the one-way authentication model does not allow $A_{k}$ to usurp the security of an individual legitimate device in the network. Let us assume that $A_{k}$ initiates an authentication request to $D_{i}$ in its close vicinity. Once $D_{i}$ receives $A_{k}$'s authentication request, it checks the MAC address of $A_{k}$ in its MAC address list. If $A_{k}$'s MAC address $\notin$ $D_{i}$ MAC-Address list, $D_{i}$ will not respond to $A_{k}$'s authentication request. In other words, $D_{i}$ will deny $A_{k}$'s authentication request in an operational network to avoid impersonation attack.   
			
			\subsubsection{Sybil Attacks}
			
			Our Hash-MAC-DSDV scheme is highly resilient  against Sybil attack, since each device has a distinct MAC address recorded in the local chain as well as in the public chain. Therefore, the usurpation of the protection of the $D {i}$ device needs to define its MAC-Address with MD5 hash function, but sensor device as an adversary has limited resources to identify the MAC address of legitimate devices by following $2^{128}$ iterations. Therefore, our model protects against Sybil attacks. 
			
			\subsubsection{Spoofing Attacks} 
			
			Spoofing attack is another disruptive assault intended to compromise the security of a network. Assume that  $A_{k}$ tries to spoof the MAC address of $D_{i}$ by launching an attack. For that, $A_{k}$ needs to know the MAC address of $D_{i}$. Likewise, $A_{k}$ needs to know the MAC address of any $D_{i}$  $\notin$  $D_{n-1}$ in the network. Consequently, $A_{k}$ will need to hijack an individual device ($CH_i$ or $D_{j}$) to get the MAC address of a legitimate device. This is not possible in the proposed model, due to D2D, local chain, and public chain authentication. The authentication request of $A_{k}$ in the proposed scheme will always be identified successfully to prevent spoofing attacks against the deployed network.   
			
			\subsubsection{Denial of Service Attack (DoS)}
			To elaborate on our scheme against DoS, assume that $A_{k}$ launches a DoS attack towards $D_{i}$ in the network. Once $D_{i}$ receives $A_{k}$'s first message request, it matches the MAC address of $A_{k}$ in its registered MAC address list. If the MAC address of $A_{k}$ $\notin$ $D_{i}$'s MAC address list, $D_i$ will deny $A_{k}$'s request and blacklist $A_{k}$ in its directory. 
			
			\subsubsection{Forward and backward secrecy}
			Hash-MAC-DSDV can offer forward and backward secrecy because the legitimate $D_{i}$ , $CH_i$ and $D_{j}$  react only to those devices listed in the local  and the public chains. These devices  first match the MAC addresses of the requesting devices in their MAC address list. Consequently, the tests of our scheme against this type of attack allow access to only legitimate devices. 
			
			\subsubsection{Base Station (BS) Impersonation Attacks}
			
			The Hash-MAC-DSDV scheme also showed effectiveness against BS impersonation attacks. Assume $A_{k}$ tries to communicate directly with BS and compromise its security.  $A_{k}$  generates communication request with the nearest $D_{j}$.  Upon reception of $A_{k}$ message request, $D_{j}$ checks the MAC address of requesting device in its local chain MAC-Address list. The MAC address of $A_{k}$ $\notin$ $D_{j}$ local chain MAC-Address list. So, $D_{j}$ will identify  $A_{k}$  successfully and will avoid BS impersonation attacks in the network. Therefore, our scheme has better results against the impersonation attack.   
			
			\begin{table}[h!]
				\caption{Statistical results analysis for different security threats}
				\begin{center}
					\begin{tabular}{  p{4.1cm}   |  p{.7cm} | p{.38cm} | p{.38cm} | p{.38cm}  }	
						\hline \textbf{ Type of Attack } & \textbf{Our Scheme} & \textbf{\cite{khalid2020decentralized}} & \textbf{\cite{gope2018lightweight}}  & \textbf{	\cite{pan2018edgechain}} \\
						\hline	Eavesdropping Attack  & Yes & Yes   & No & Yes   \\
						\hline	Sensing Device Impersonate Attack  & Yes & Yes   & Yes & Yes  \\
						\hline	Sybil Attacks  & Yes & No    & Yes & No  \\
						\hline	Spoofing Attacks  & Yes & Yes    & No & Yes  \\	
						\hline	Denial of Service Attack (DoS) & Yes & Yes   & Yes  & No  \\
						\hline	Spoofing Attacks  & Yes & Yes  & Yes & Yes \\
						\hline		Perfect forward and backward secrecy  & Yes & Yes    & No & Yes \\
						\hline	Base Station (BS) impersonate Attacks  & Yes & Yes    & Yes & Yes \\
						
						\hline
					\end{tabular}
				\end{center}
				\label{tb2}
			\end{table}
			
			\subsection{Computation Cost Comparative Analysis }
			
			The computation results can easily be evaluated from the proposed model execution time in terms of processing, energy consumption and memory utilization. The operation of the model, furthermore, depicts that most of the computation is performed at the cluster head and base station, which have higher processing power. Similarly, upon registration of MAC addresses at the base station $D_j$, each $D_j$ advertises the information in the public chain followed by CH nodes in the local chain. Consequently, the legitimate devices in the local chain update their routing table according to CH advertised information. The legitimate devices follow routing table information to transmit their collected data in the network. However, this is a one-way process, because the legitimate devices transmit their data in a unicast  fashion following their routing table information, which not only minimizes the energy consumption, but also minimizes the calculation or processing of legitimate devices and as a whole, improves the network performance. Therefore, the computation cost of the proposed model is better than the existing schemes of \cite{khalid2020decentralized}, \cite{gope2018lightweight}, and \cite{pan2018edgechain}. Khalid et al. \cite{khalid2020decentralized} scheme has high computation because the next hop update  is the responsibility of an individual participating device. Therefore, it consumes more energy with network overhead and higher computation cost. Likewise, Gope et al. \cite{gope2018lightweight}, scheme has also higher computation cost in comparison to our scheme, because they used two factor authentication model in their scheme with physical assessment of IoT devices. Moreover, the participating devices of the network updated their routing table after define interval of time, which also creates contention with higher computation cost. The complex model implementation of  \cite{pan2018edgechain} increases its computation cost in an operational environment. 
			
			\subsection{Communication Cost Comparative Analysis}
			
			The communication cost is another important aspect to consider while designing a new protocol or modifying existing protocols. Therefore, the communication cost of any IoT network determines its capacity, performance, and reliability. The communication cost of the proposed model shall be assessed with a pricing structure based on the following statement:
			
			\begin{enumerate}
				\item The identity of legitimate device as its MAC address
				\item MD5 Hash function with message digest
				\item Time stamp for key sharing via CH and base station
				\item Authentication procedure 
			\end{enumerate}         
			
			The simple registration and authentication mechanism of legitimate devices in the proposed model minimizes the communication costs up-to significant level in the presence of rivals schemes. Moreover, the rival schemes of \cite{khalid2020decentralized}, \cite{gope2018lightweight}, and \cite{pan2018edgechain}, use an authentication process between BS/Edge devices or device to exchange information in the network, every time. Keeping in view, the authentication procedure adopted by the existing schemes, our proposed model has the simplest authentication process with minimal resource consumption and better results.
			
			\subsection{Comparative Analysis on the Accuracy of Threat Detection}
			
\begin{figure} [h!]
			\centering
				\includegraphics[width=0.8\linewidth, height = 4.5cm]{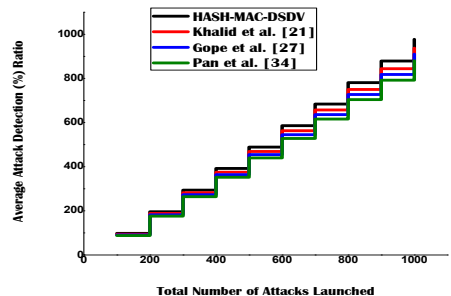}
				\caption{Comparative statistical analysis of our scheme with rival schemes for threat detections}
				\label{fig : 4}
			\end{figure}

\begin{figure} [h!]
			\centering
				\includegraphics[width=0.8\linewidth, height = 4.5cm]{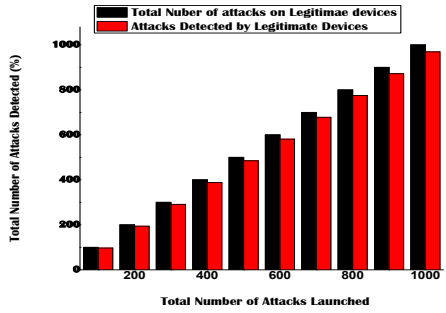}
				\caption{Statistical analysis of legitimate devices to detect malicious attacks in an operational network}
				\label{fig : 5}
			\end{figure}

\begin{figure} [h!]
			\centering
				\includegraphics[width=0.8\linewidth, height = 4.5cm]{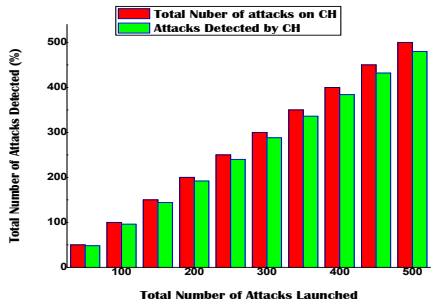}
				\caption{Statistical analysis of CH node to detect malicious attacks in an operational network}
				\label{fig : 6}
			\end{figure}

\begin{figure} [h!]
			\centering
				\includegraphics[width=0.8\linewidth, height = 4.5cm]{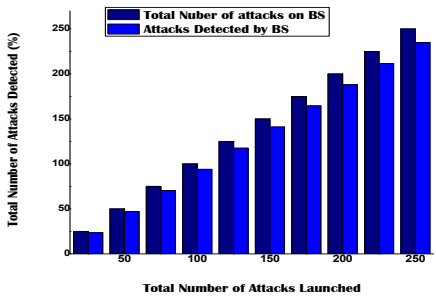}
				\caption{Statistical analysis of BS to detect malicious attacks in an operational network}
				\label{fig : 7}
			\end{figure}

			The proposed model was assessed with its competitor schemes based on accurate threat detection in an operational network. The Hash-MAC-DSDV scheme's utmost objective is to detect and report malicious activity in the deployed network. Although the competitor schemes resolve the legitimate device authentication issue to some extent, they were flawed in addressing issues like D2D, device-to-CH, device-to-BS, and CH-to-BS authentication at the same time. In our scheme's simulation environment, we have evaluated the proposed model for aforementioned attacks. Moreover, we have also launched attacks on legitimate devices, CH(s) and BS(s) to verify the effectiveness of our scheme. The detection rate of malicious device in the proposed model was 98.2\%, which showed an average 15\% improvement over the existing schemes. Likewise, during simulation analysis, we have changed the number of malicious devices, fake packets, target area such as participating devices, CH nodes, and BS(s) to overview our scheme's reliability.
			Overall, the performance of the proposed model was significant compared to existing rival schemes to combat malicious attacks in IoT networks connected in the form of multi-WSNs. Results for our scheme and other schemes are presented in Figure~\ref{fig : 4}. In addition, in Figures~\ref{fig : 5},~\ref{fig : 6}, and~\ref{fig : 7}, the results for individual network components such as devices, CH and BS against malicious activity in an operating network are presented.                       
			
			
			\subsection{Energy Consumption}
			
\begin{figure} [h!]
			\centering
				\includegraphics[width=0.8\linewidth, height = 4.5cm]{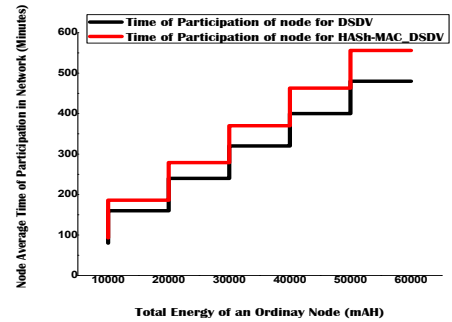}
				\caption{Energy consumption statistical analysis of Hash-MAC-DSDV and ordinary DSDV protocol}
				\label{fig : 8}
			\end{figure}

			Sensor devices are sensitive and have limited resources memory, and energy, therefore, efficient utilization of these devices increases its productivity in terms of network lifespan. Therefore, while designing Hash-MAC-DSDV mutual authentication scheme for multi-WSNs connected in the form of IoT network, we considered limited resources of sensor devices. The proposed model's energy consumption was evaluated with ordinary DSDV protocol in simulation environment. The results statistics captured during Hash-MAC-DSDV protocol simulation showed improvement in the lifetime of ordinary devices working in the network over the ordinary DSDV protocol. Likewise, while evaluating DSDV protocol, participating devices consume more energy as compared to our scheme, due to continued route update with neighbor devices. Conversely, in our scheme, the legitimate device updates its routing table to local and public chain information to exchange data in the network. Therefore, device participation in our scheme surpasses ordinary DSDV protocol by 11\% improvement in an operational network as observed during simulation analysis. The results of both routing protocols are shown in Figure~\ref{fig : 8}.       
			
			\subsection{Latency Results Statistical Analysis}
			
\begin{figure} [h!]
			\centering
				\includegraphics[width=0.8\linewidth, height = 4.5cm]{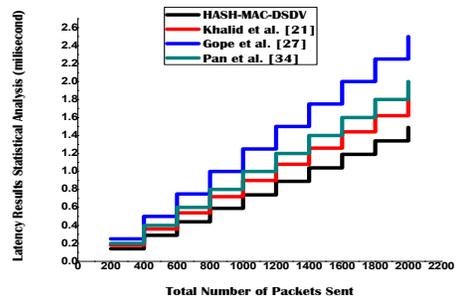}
				\caption{Latency Results statistical analysis of Hash-MAC-DSDV scheme with its rival schemes}
				\label{fig : 9}
			\end{figure}

			Latency is an important aspect of wireless networks when designing protocols. We therefore considered consistency in time during the design stage of our scheme in order to exchange information effectively throughout the network. In the simulation environment, the number of devices and local chains is gradually increased to overview the duration of the exchange of messages in the network. Subsequently, malicious devices were introduced in the network with a fake authentication request in the local chain as well as in the public chain to observe the latency of legitimate device messages. Although, during the presence of malicious devices, network traffic was at its peak, the legitimate device showed consistency while exchanging data on the network. The unique pattern of communication among the legitimate devices in the proposed model played a vital role in ensuring time-consistency during the exchange of messages. The results of the proposed scheme are compared with competing schemes and are presented in Figure~\ref{fig : 9}.
			
			\subsection{Packet lost Ratio Statistical Results Analysis}
			Wireless communication is susceptible to various attacks and environmental factors. Therefore, packet lost ratio is given preference to define the performance reliability of any wireless network. During simulation of Hash-MAC-DSDV protocol the PLR was quantified through the following formula: 
			\begin{equation}
			PLR = Packet \ sent \ - \ Packet \ received 
			\end{equation}     
			
\begin{figure} [h!]
			\centering
				\includegraphics[width=0.8\linewidth, height = 4.5cm]{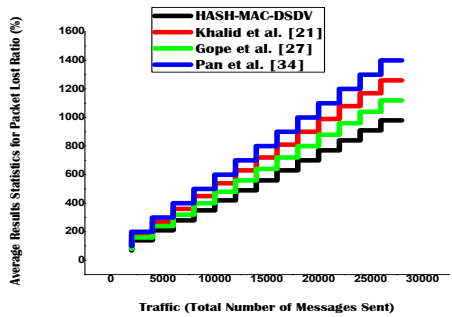}
				\caption{PLR Results statistical analysis for Hash-MAC-DSDV scheme with competitors schemes}
				\label{fig : 10}
			\end{figure}

			The network traffic was increased in simulation and we quantified the ratio between packets sent and received, which showed reliable results during analysis. Moreover, the consistency of PLR was verified by targeting devices during traffic congestion in an operational network. Our results showed only 7 \% PLR for the proposed model during peak traffic, verifying the significance of our scheme over the compared schemes. The statistical analysis for PLR is shown in Figure~\ref{fig : 10}. 
			
			\section{Conclusion}
			
			In this paper, we proposed Hash-MAC-DSDV, a mutual authentication scheme for CPS connected to form an IoT network. We modified the DSDV protocol to enable mutual authentication among participating devices via hop count communication. Initially, all legitimate devices  registered their MAC addresses with the base stations via concerned cluster heads to form local and public chains, respectively. Upon  registration, the base station applies MD5 hash algorithm to the registered MAC addresses for advertisement in the public chain, where a number of cluster heads are involved. These cluster heads advertised the base station information in the local chain and the devices update their routing table according to the advertised information. Likewise, each legitimate device uses its routing table information to send data to a destination. However, to verify the legitimacy of adjacent devices or cluster head during hop-count authentication, matching of MAC addresses is followed by each device.  Our proposed scheme outperforms existing approaches in terms of attack detection rate, computation cost, communication cost, energy consumption, PLR, and latency. The unicast communication and one-way-hash authentication of our model increases its applicability in real deployments with minimal resource consumption overhead.     
			
			In the future, we are looking to implement the proposed model in the real environment, where a number of local and public chains network should be interconnect to verify the performance reliability and consistency ratio.  
			\section*{Acknowledgment}
			This work is partially supported by the NIH (P20GM109090), NSF (CNS-2016714), the Nebraska University Collaboration Initiative, and the Nebraska Tobacco Settlement Biomedical Research Development Funds. This research is also partially funded by the Deanship of Scientific Research at Princess Nourah Bint Abdul Rahman University through the Fast-track Research Funding Program.
			\bibliographystyle{unsrt}
			\bibliography{library}

\begin{thebibliography}{10}

\bibitem{ding2019survey}
Wenxiu Ding, Xuyang Jing, Zheng Yan, and Laurence~T Yang.
\newblock A survey on data fusion in internet of things: Towards secure and
  privacy-preserving fusion.
\newblock {\em Information Fusion}, 51:129--144, 2019.

\bibitem{jiang2019user}
Qi~Jiang, Yuanyuan Qian, Jianfeng Ma, Xindi Ma, Qingfeng Cheng, and Fushan Wei.
\newblock User centric three-factor authentication protocol for cloud-assisted
  wearable devices.
\newblock {\em International Journal of Communication Systems}, 32(6):e3900,
  2019.

\bibitem{wang2019energy}
Quan Wang, Deyu Lin, Pengfei Yang, and Zhiqiang Zhang.
\newblock An energy-efficient compressive sensing-based clustering routing
  protocol for wsns.
\newblock {\em IEEE Sensors Journal}, 19(10):3950--3960, 2019.

\bibitem{gubbi2013internet}
Jayavardhana Gubbi, Rajkumar Buyya, Slaven Marusic, and Marimuthu Palaniswami.
\newblock Internet of things (iot): A vision, architectural elements, and
  future directions.
\newblock {\em Future generation computer systems}, 29(7):1645--1660, 2013.

\bibitem{khan2018iot}
Minhaj~Ahmad Khan and Khaled Salah.
\newblock Iot security: Review, blockchain solutions, and open challenges.
\newblock {\em Future Generation Computer Systems}, 82:395--411, 2018.

\bibitem{mahmod2020robust}
Md~Jubayer~al Mahmod and Ujjwal Guin.
\newblock A robust, low-cost and secure authentication scheme for iot
  applications.
\newblock {\em Cryptography}, 4(1):8, 2020.

\bibitem{ferrag2017privacy}
Mohamed~Amine Ferrag, Leandros Maglaras, and Ahmed Ahmim.
\newblock Privacy-preserving schemes for ad hoc social networks: A survey.
\newblock {\em IEEE Communications Surveys \& Tutorials}, 19(4):3015--3045,
  2017.

\bibitem{huang2019towards}
Junqin Huang, Linghe Kong, Guihai Chen, Min-You Wu, Xue Liu, and Peng Zeng.
\newblock Towards secure industrial iot: Blockchain system with credit-based
  consensus mechanism.
\newblock {\em IEEE Transactions on Industrial Informatics}, 15(6):3680--3689,
  2019.

\bibitem{biswas2018scalable}
Sujit Biswas, Kashif Sharif, Fan Li, Boubakr Nour, and Yu~Wang.
\newblock A scalable blockchain framework for secure transactions in iot.
\newblock {\em IEEE Internet of Things Journal}, 6(3):4650--4659, 2018.

\bibitem{wu2018lightweight}
Fan Wu, Xiong Li, Arun~Kumar Sangaiah, Lili Xu, Saru Kumari, Liuxi Wu, and Jian
  Shen.
\newblock A lightweight and robust two-factor authentication scheme for
  personalized healthcare systems using wireless medical sensor networks.
\newblock {\em Future Generation Computer Systems}, 82:727--737, 2018.

\bibitem{hammi2018bubbles}
Mohamed~Tahar Hammi, Badis Hammi, Patrick Bellot, and Ahmed Serhrouchni.
\newblock Bubbles of trust: A decentralized blockchain-based authentication
  system for iot.
\newblock {\em Computers \& Security}, 78:126--142, 2018.

\bibitem{yu2019wireless}
A~Yu, Jia-Jia Ji, Yuan Wang, Hong-Bing Sun, et~al.
\newblock Wireless monitoring system for corrosion degree of reinforcement in
  concrete.
\newblock {\em Journal of Nanoelectronics and Optoelectronics}, 14(6):887--893,
  2019.

\bibitem{adil2020anonymous}
Muhammad Adil, Mohammed~Amin Almaiah, Alhuseen Omar~Alsayed, and Omar Almomani.
\newblock An anonymous channel categorization scheme of edge nodes to detect
  jamming attacks in wireless sensor networks.
\newblock {\em Sensors}, 20(8):2311, 2020.

\bibitem{khalid2020decentralized}
Umair Khalid, Muhammad Asim, Thar Baker, Patrick~CK Hung, Muhammad~Adnan Tariq,
  and Laura Rafferty.
\newblock A decentralized lightweight blockchain-based authentication mechanism
  for iot systems.
\newblock {\em Cluster Computing}, pages 1--21, 2020.

\bibitem{tonyali2018privacy}
Samet Tonyali, Kemal Akkaya, Nico Saputro, A~Selcuk Uluagac, and Mehrdad
  Nojoumian.
\newblock Privacy-preserving protocols for secure and reliable data aggregation
  in iot-enabled smart metering systems.
\newblock {\em Future Generation Computer Systems}, 78:547--557, 2018.

\bibitem{aloqaily2019data}
Moayad Aloqaily, Ismaeel Al~Ridhawi, Haythem~Bany Salameh, and Yaser Jararweh.
\newblock Data and service management in densely crowded environments:
  Challenges, opportunities, and recent developments.
\newblock {\em IEEE Communications Magazine}, 57(4):81--87, 2019.

\bibitem{baker2020secure}
Thar Baker, Muhammad Asim, {\'A}ine MacDermott, Farkhund Iqbal, Faouzi Kamoun,
  Babar Shah, Omar Alfandi, and Mohammad Hammoudeh.
\newblock A secure fog-based platform for scada-based iot critical
  infrastructure.
\newblock {\em Software: Practice and Experience}, 50(5):503--518, 2020.

\bibitem{rathee2019blockchain}
Geetanjali Rathee, Ashutosh Sharma, Razi Iqbal, Moayad Aloqaily, Naveen Jaglan,
  and Rajiv Kumar.
\newblock A blockchain framework for securing connected and autonomous
  vehicles.
\newblock {\em Sensors}, 19(14):3165, 2019.

\bibitem{tariq2019security}
Noshina Tariq, Muhammad Asim, Feras Al-Obeidat, Muhammad Zubair~Farooqi, Thar
  Baker, Mohammad Hammoudeh, and Ibrahim Ghafir.
\newblock The security of big data in fog-enabled iot applications including
  blockchain: a survey.
\newblock {\em Sensors}, 19(8):1788, 2019.

\bibitem{gope2018lightweight}
Prosanta Gope and Biplab Sikdar.
\newblock Lightweight and privacy-preserving two-factor authentication scheme
  for iot devices.
\newblock {\em IEEE Internet of Things Journal}, 6(1):580--589, 2018.

\bibitem{feng2018aaot}
Wei Feng, Yu~Qin, Shijun Zhao, and Dengguo Feng.
\newblock Aaot: Lightweight attestation and authentication of low-resource
  things in iot and cps.
\newblock {\em Computer Networks}, 134:167--182, 2018.

\bibitem{li2018creditcoin}
Lun Li, Jiqiang Liu, Lichen Cheng, Shuo Qiu, Wei Wang, Xiangliang Zhang, and
  Zonghua Zhang.
\newblock Creditcoin: A privacy-preserving blockchain-based incentive
  announcement network for communications of smart vehicles.
\newblock {\em IEEE Transactions on Intelligent Transportation Systems},
  19(7):2204--2220, 2018.

\bibitem{cui2018detection}
Zhihua Cui, Fei Xue, Xingjuan Cai, Yang Cao, Gai-ge Wang, and Jinjun Chen.
\newblock Detection of malicious code variants based on deep learning.
\newblock {\em IEEE Transactions on Industrial Informatics}, 14(7):3187--3196,
  2018.

\bibitem{aitzhan2016security}
Nurzhan~Zhumabekuly Aitzhan and Davor Svetinovic.
\newblock Security and privacy in decentralized energy trading through
  multi-signatures, blockchain and anonymous messaging streams.
\newblock {\em IEEE Transactions on Dependable and Secure Computing},
  15(5):840--852, 2016.

\bibitem{salman2018security}
Tara Salman, Maede Zolanvari, Aiman Erbad, Raj Jain, and Mohammed Samaka.
\newblock Security services using blockchains: A state of the art survey.
\newblock {\em IEEE Communications Surveys \& Tutorials}, 21(1):858--880, 2018.

\bibitem{cai2019ensemble}
Xingjuan Cai, Jiangjiang Zhang, Hao Liang, Lei Wang, and Qidi Wu.
\newblock An ensemble bat algorithm for large-scale optimization.
\newblock {\em International Journal of Machine Learning and Cybernetics},
  10(11):3099--3113, 2019.

\bibitem{pan2018edgechain}
Jianli Pan, Jianyu Wang, Austin Hester, Ismail Alqerm, Yuanni Liu, and Ying
  Zhao.
\newblock Edgechain: An edge-iot framework and prototype based on blockchain
  and smart contracts.
\newblock {\em IEEE Internet of Things Journal}, 6(3):4719--4732, 2018.

\bibitem{bao2018iotchain}
Zijian Bao, Wenbo Shi, Debiao He, and Kim-Kwang~Raymond Chood.
\newblock Iotchain: A three-tier blockchain-based iot security architecture.
\newblock {\em arXiv preprint arXiv:1806.02008}, 2018.

\bibitem{won2017certificateless}
Jongho Won, Seung-Hyun Seo, and Elisa Bertino.
\newblock Certificateless cryptographic protocols for efficient drone-based
  smart city applications.
\newblock {\em IEEE Access}, 5:3721--3749, 2017.

\end{thebibliography}


\begin{thebibliography}{10}

\bibitem{gubbi2013internet}
Jayavardhana Gubbi, Rajkumar Buyya, Slaven Marusic, and Marimuthu Palaniswami.
\newblock Internet of things (iot): A vision, architectural elements, and
  future directions.
\newblock {\em Future generation computer systems}, 29(7):1645--1660, 2013.

\bibitem{khan2020secured}
Fazlullah Khan, Mian~Ahmad Jan, Ateeq~Ur Rehman, Spyridon Mastorakis, Mamoun
  Alazab, and Paul Watters.
\newblock A secured and intelligent communication scheme for iiot-enabled
  pervasive edge computing.
\newblock {\em IEEE Transactions on Industrial Informatics}, 2020.

\bibitem{jan2020security}
Mian~Ahmad Jan, Jinjin Cai, Xiang-Chuan Gao, Fazlullah Khan, Spyridon
  Mastorakis, Muhammad Usman, Mamoun Alazab, and Paul Watters.
\newblock Security and blockchain convergence with internet of multimedia
  things: Current trends, research challenges and future directions.
\newblock {\em Journal of Network and Computer Applications}, page 102918,
  2020.

\bibitem{mastorakis2020icedge}
Spyridon Mastorakis, Abderrahmen Mtibaa, Jonathan Lee, and Satyajayant Misra.
\newblock Icedge: When edge computing meets information-centric networking.
\newblock {\em IEEE Internet of Things Journal}, 7(5):4203--4217, 2020.

\bibitem{khan2018iot}
Minhaj~Ahmad Khan and Khaled Salah.
\newblock Iot security: Review, blockchain solutions, and open challenges.
\newblock {\em Future Generation Computer Systems}, 82:395--411, 2018.

\bibitem{zhong2020automated}
Xin Zhong, Pei-Chi Huang, Spyridon Mastorakis, and Frank~Y Shih.
\newblock An automated and robust image watermarking scheme based on deep
  neural networks.
\newblock {\em arXiv preprint arXiv:2007.02460}, 2020.

\bibitem{jan2020lightweight}
Mian~Ahmad Jan, Fazlullah Khan, Rahim Khan, Spyridon Mastorakis, Varun~G Menon,
  Paul Watters, and Mamoun Alazab.
\newblock A lightweight mutual authentication and privacy-preservation scheme
  for intelligent wearable devices in industrial-cps.
\newblock {\em IEEE Transactions on Industrial Informatics}, 2020.

\bibitem{shannigrahi2020next}
Susmit Shannigrahi, Spyridon Mastorakis, and Francisco~R Ortega.
\newblock Next-generation networking and edge computing for mixed reality
  real-time interactive systems.
\newblock In {\em 2020 IEEE International Conference on Communications
  Workshops (ICC Workshops)}, pages 1--6. IEEE, 2020.

\bibitem{mahmod2020robust}
Md~Jubayer~al Mahmod and Ujjwal Guin.
\newblock A robust, low-cost and secure authentication scheme for iot
  applications.
\newblock {\em Cryptography}, 4(1):8, 2020.

\bibitem{ferrag2017privacy}
Mohamed~Amine Ferrag, Leandros Maglaras, and Ahmed Ahmim.
\newblock Privacy-preserving schemes for ad hoc social networks: A survey.
\newblock {\em IEEE Communications Surveys \& Tutorials}, 19(4):3015--3045,
  2017.

\bibitem{ullah2020icn}
Rehmat Ullah, Muhammad Atif~Ur Rehman, Muhammad~Ali Naeem, Byung-Seo Kim, and
  Spyridon Mastorakis.
\newblock Icn with edge for 5g: Exploiting in-network caching in icn-based edge
  computing for 5g networks.
\newblock {\em Future Generation Computer Systems}, 111:159--174, 2020.

\bibitem{mastorakis2020dapes}
Spyridon Mastorakis, Tianxiang Li, and Lixia Zhang.
\newblock {DAPES: Named Data for Off-the-Grid File Sharing with Peer-to-Peer
  Interactions}.
\newblock {\em 40th IEEE International Conference on Distributed Computing
  Systems (ICDCS)}, 2020.

\bibitem{rehman2020ccic}
Muhammad Atif~Ur Rehman, Rehmat Ullah, Byung-Seo Kim, Boubakr Nour, and
  Spyridon Mastorakis.
\newblock Ccic-wsn: An architecture for single channel cluster-based
  information-centric wireless sensor networks.
\newblock {\em IEEE Internet of Things Journal}, 2020.

\bibitem{li2019distributed}
Tianxiang Li, Zhaoning Kong, Spyridon Mastorakis, and Lixia Zhang.
\newblock Distributed dataset synchronization in disruptive networks.
\newblock In {\em 2019 IEEE 16th International Conference on Mobile Ad Hoc and
  Sensor Systems (MASS)}, pages 428--437. IEEE, 2019.

\bibitem{yu2019wireless}
A~Yu, Jia-Jia Ji, Yuan Wang, Hong-Bing Sun, et~al.
\newblock Wireless monitoring system for corrosion degree of reinforcement in
  concrete.
\newblock {\em Journal of Nanoelectronics and Optoelectronics}, 14(6):887--893,
  2019.

\bibitem{adil2020anonymous}
Muhammad Adil, Mohammed~Amin Almaiah, Alhuseen Omar~Alsayed, and Omar Almomani.
\newblock An anonymous channel categorization scheme of edge nodes to detect
  jamming attacks in wireless sensor networks.
\newblock {\em Sensors}, 20(8):2311, 2020.

\bibitem{khalid2020decentralized}
Umair Khalid, Muhammad Asim, Thar Baker, Patrick~CK Hung, Muhammad~Adnan Tariq,
  and Laura Rafferty.
\newblock A decentralized lightweight blockchain-based authentication mechanism
  for iot systems.
\newblock {\em Cluster Computing}, pages 1--21, 2020.

\bibitem{tonyali2018privacy}
Samet Tonyali, Kemal Akkaya, Nico Saputro, A~Selcuk Uluagac, and Mehrdad
  Nojoumian.
\newblock Privacy-preserving protocols for secure and reliable data aggregation
  in iot-enabled smart metering systems.
\newblock {\em Future Generation Computer Systems}, 78:547--557, 2018.

\bibitem{aloqaily2019data}
Moayad Aloqaily, Ismaeel Al~Ridhawi, Haythem~Bany Salameh, and Yaser Jararweh.
\newblock Data and service management in densely crowded environments:
  Challenges, opportunities, and recent developments.
\newblock {\em IEEE Communications Magazine}, 57(4):81--87, 2019.

\bibitem{baker2020secure}
Thar Baker, Muhammad Asim, {\'A}ine MacDermott, Farkhund Iqbal, Faouzi Kamoun,
  Babar Shah, Omar Alfandi, and Mohammad Hammoudeh.
\newblock A secure fog-based platform for scada-based iot critical
  infrastructure.
\newblock {\em Software: Practice and Experience}, 50(5):503--518, 2020.

\bibitem{rathee2019blockchain}
Geetanjali Rathee, Ashutosh Sharma, Razi Iqbal, Moayad Aloqaily, Naveen Jaglan,
  and Rajiv Kumar.
\newblock A blockchain framework for securing connected and autonomous
  vehicles.
\newblock {\em Sensors}, 19(14):3165, 2019.

\bibitem{tariq2019security}
Noshina Tariq, Muhammad Asim, Feras Al-Obeidat, Muhammad Zubair~Farooqi, Thar
  Baker, Mohammad Hammoudeh, and Ibrahim Ghafir.
\newblock The security of big data in fog-enabled iot applications including
  blockchain: a survey.
\newblock {\em Sensors}, 19(8):1788, 2019.

\bibitem{gope2018lightweight}
Prosanta Gope and Biplab Sikdar.
\newblock Lightweight and privacy-preserving two-factor authentication scheme
  for iot devices.
\newblock {\em IEEE Internet of Things Journal}, 6(1):580--589, 2018.

\bibitem{feng2018aaot}
Wei Feng, Yu~Qin, Shijun Zhao, and Dengguo Feng.
\newblock Aaot: Lightweight attestation and authentication of low-resource
  things in iot and cps.
\newblock {\em Computer Networks}, 134:167--182, 2018.

\bibitem{li2018creditcoin}
Lun Li, Jiqiang Liu, Lichen Cheng, Shuo Qiu, Wei Wang, Xiangliang Zhang, and
  Zonghua Zhang.
\newblock Creditcoin: A privacy-preserving blockchain-based incentive
  announcement network for communications of smart vehicles.
\newblock {\em IEEE Transactions on Intelligent Transportation Systems},
  19(7):2204--2220, 2018.

\bibitem{cui2018detection}
Zhihua Cui, Fei Xue, Xingjuan Cai, Yang Cao, Gai-ge Wang, and Jinjun Chen.
\newblock Detection of malicious code variants based on deep learning.
\newblock {\em IEEE Transactions on Industrial Informatics}, 14(7):3187--3196,
  2018.

\bibitem{aitzhan2016security}
Nurzhan~Zhumabekuly Aitzhan and Davor Svetinovic.
\newblock Security and privacy in decentralized energy trading through
  multi-signatures, blockchain and anonymous messaging streams.
\newblock {\em IEEE Transactions on Dependable and Secure Computing},
  15(5):840--852, 2016.

\bibitem{salman2018security}
Tara Salman, Maede Zolanvari, Aiman Erbad, Raj Jain, and Mohammed Samaka.
\newblock Security services using blockchains: A state of the art survey.
\newblock {\em IEEE Communications Surveys \& Tutorials}, 21(1):858--880, 2018.

\bibitem{cai2019ensemble}
Xingjuan Cai, Jiangjiang Zhang, Hao Liang, Lei Wang, and Qidi Wu.
\newblock An ensemble bat algorithm for large-scale optimization.
\newblock {\em International Journal of Machine Learning and Cybernetics},
  10(11):3099--3113, 2019.

\bibitem{pan2018edgechain}
Jianli Pan, Jianyu Wang, Austin Hester, Ismail Alqerm, Yuanni Liu, and Ying
  Zhao.
\newblock Edgechain: An edge-iot framework and prototype based on blockchain
  and smart contracts.
\newblock {\em IEEE Internet of Things Journal}, 6(3):4719--4732, 2018.

\bibitem{bao2018iotchain}
Zijian Bao, Wenbo Shi, Debiao He, and Kim-Kwang~Raymond Chood.
\newblock Iotchain: A three-tier blockchain-based iot security architecture.
\newblock {\em arXiv preprint arXiv:1806.02008}, 2018.

\bibitem{won2017certificateless}
Jongho Won, Seung-Hyun Seo, and Elisa Bertino.
\newblock Certificateless cryptographic protocols for efficient drone-based
  smart city applications.
\newblock {\em IEEE Access}, 5:3721--3749, 2017.

\end{thebibliography}

		    \section*{Biographies}
		    
		    \vskip -5.0\baselineskip plus -1fil
			
			\begin{IEEEbiography}[{\includegraphics[width=1in,height=1.25in,clip,keepaspectratio]{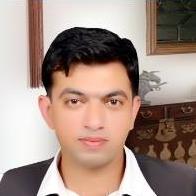}}]{\textbf{Muhammad Adil}} received his Associate Engineer degree in Electronics form the school of Electronic associated with civil aviation Pakistan in 2010. Mr. Adil received his Bachelor of Science in Computer Science (4 years programs) and Master of Science in Computer Sciences (2 years program) with specialization in Computer Networks from Virtual University of Lahore, Pakistan in 2016 and 2019, respectively. He has CCNA and CCNP certification. He is currently a PhD Candidate. His research area includes different routing protocols, Security, and Load Balancing in WSN, IoT, and ad hoc networks. Moreover, Mr. Adil is also interested in Dynamic Wireless Charging of Electric Vehicles connected in network topological infrastructure with Machine learning techniques. He has many publications in prestigious journals such as IEEE Access, IEEE Sensors, Computer Networks Elsevier, CMC-Computer Material \& Continua and MDPI Sensor etc. In addition, he is IEEE Student member. He is reviewing for prestigious journals, such as IEEE Access, IEEE Sensors, IEEE Systems, IEEE Internet of Things, IEEE Transaction of Industrial Informatics, IEEE Transactions on Cognitive Communications and Networking, MDPI Sensors and Computer Networks Elsevier Journals, Telecommunication System, and IEEE Wireless Communication Letters.
			\end{IEEEbiography}
			
			\vskip -2.0\baselineskip plus -1fil
			
			\begin{IEEEbiography}[{\includegraphics[width=1in,height=1.25in,clip,keepaspectratio]{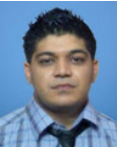}}]{MIAN AHMAD JAN}   (Senior Member, IEEE) received the Ph.D. degree in computer systems
				from the University of Technology Sydney (UTS),	Australia, in 2016. He is an Assistant Professor with Abdul Wai Khan University Mardan, Pakistan. His research articles have been published in various prestigious IEEE TRANSACTIONS and	journals (Elsevier). His research interests include security and privacy in the Internet of Things and wireless sensor networks. He had been the recipient of various prestigious scholarships during his studies, notably the	International Research Scholarship (IRS) from the UTS, Australia, and the Commonwealth Scientific Industrial Research Organization (CSIRO) Scholarships. He was awarded the Best Researcher Award from the UTS in 2014. He was the General Co-Chair of Springer/EAI Second International Conference on Future Intelligent Vehicular Technologies in 2017. He has	been a Guest Editor of numerous special issues in various prestigious journals such as Future Generation Computing Systems (Elsevier), Mobile Networks and Applications (MONET) (Springer), Ad Hoc and Sensor Wireless Networks, and Information (MDPI).
			\end{IEEEbiography}
			
			\vskip -2.0\baselineskip plus -1fil
			
			\begin{IEEEbiography}[{\includegraphics[width=1in,height=1.25in,clip,keepaspectratio]{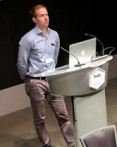}}]{Spyridon Mastorakis}  (Member, IEEE) is an Assistant Professor in Computer Science at the University of Nebraska Omaha. He received his Ph.D. in Computer Science from the University of California, Los Angeles (UCLA) in 2019. He also received an M.S. in Computer Science from UCLA in 2017 and a 5-year diploma (equivalent to M.Eng.) in Electrical and Computer Engineering from the National Technical University of Athens (NTUA) in 2014. His research interests include network systems and protocols, Internet architectures (such as Information-Centric Networking and Named-Data Networking), edge computing, and IoT.
			\end{IEEEbiography}
			
			\vskip -2.0\baselineskip plus -1fil
			
			\begin{IEEEbiography}[{\includegraphics[width=1in,height=1.25in,clip,keepaspectratio]{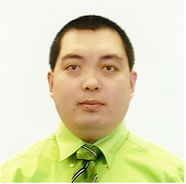}}]{\textbf{Houbing Song}} (Senior Member, IEEE) received the M.S. degree in civil engineering from the University of Texas, El Paso, TX, USA, in December 2006, and the Ph.D. degree in electrical engineering from the University of Virginia, Charlottesville, VA, USA, in August 2012.	In August 2017, he joined the Department of Electrical, Computer, Software, and Systems Engineering, Embry–Riddle Aeronautical University, Daytona Beach, FL, USA, where he is currently an Assistant Professor and the Director of the Security and Optimization for Networked Globe Laboratory. His research has been featured by popular news media outlets, including IEEE GlobalSpec’s Engineering360,	USA Today, U.S. News \& World Report, Fox News, the Association for Unmanned Vehicle Systems International, Forbes, WFTV, and New Atlas. His research interests include cyber–physical systems, cyber security and privacy, Internet of Things, edge computing, AI/machine learning, big data analytics, unmanned aircraft systems, connected vehicle, smart and connected health, and wireless communications and networking.
			\end{IEEEbiography}
			
			\vskip -2.0\baselineskip plus -1fil
			
			\begin{IEEEbiography}[{\includegraphics[width=1in,height=1.25in,clip,keepaspectratio]{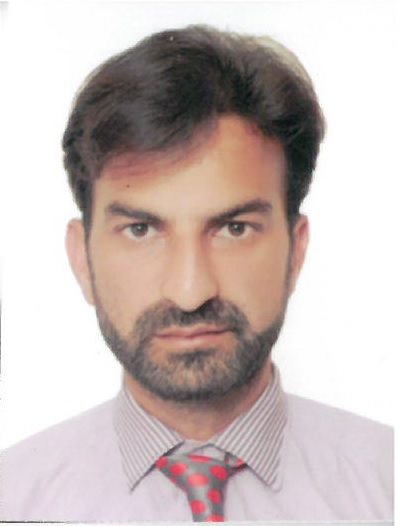}}]{ MUHAMMAD MOHSIN JADOON} received his B.S. and M.S. degrees in Electronic engineering
				from COMSATS University Islamabad, and International Islamic University Islamabad (IIUI),Pakistan, in 2007 and 2011, respectively. Dr. M Mohsin Jadoon did PhD in Split degree programs i.e. course work from IIUI and research from Queen Merry University (QMU) London, UK in 2018. Currently, he is post-doctorate research fellow at Department of Radiology and Imaging processing, Yale university, New Haven, CT, USA. He is also Lecturer with the Electrical Engineering Department, International Islamic University Islamabad.	His research interests include  Signal \& Processing, Sensors and biomedical Imaging.
			\end{IEEEbiography}
			
			\vskip -2.0\baselineskip plus -1fil
			
			\begin{IEEEbiography}[{\includegraphics[width=1in,height=1.25in,clip,keepaspectratio]{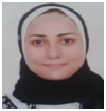}}]{Safia Abbas}
				works as associate professor with the Department of Computer Science, Faculty of Computer and Information Sciences, Princess NourahbintAbdulrahman University, KSA, during 2019-2021, and University of Ain Shams, Cairo, Egypt during 2016-2018. During 2006-2011, she received the Ph.D. from the Graduate School of Science and Technology, Niigata University, Japan. A strong theme of her work is in the swarm optimizers, and security in cloud, Medical Diagnosis using machine learning and Data mining.
			\end{IEEEbiography}
			
			\vskip -2.0\baselineskip plus -1fil
			
			\begin{IEEEbiography}[{\includegraphics[width=1in,height=1.25in,clip,keepaspectratio]{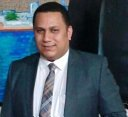}}]{Ahmed Farouk} (Member, IEEE) is currently assistant professor, before that he was a Postdoctoral Research Fellow at Wilfrid Laurier University and Ryerson University, Canada. He received his M.Sc. and Ph.D. degrees from Mansoura University, Egypt. He is one of the Top 20 technical co-founders of the Quantum Machine Learning Program by Creative Destruction Lab at the University of Toronto. Furthermore, he is selected as Top 25 of Innovate TO 150 Canada to showcase the best of Toronto’s next generation of change-makers, innovators, and entrepreneurs. He is exceptionally well known for his seminal contributions to theories of Quantum Information, Communication, and Cryptography. He published 62 papers in reputed and high impact journals like Nature Scientific Reports and Physical Review A. The exceptional quality of his research is recognized nationally and internationally.  He selected by the scientific review panel of the Council for the Lindau Nobel Laureate Meetings to participate in the 70th Lindau Nobel Laureate Meeting. His volunteering work is apparent since he appointed as chair of the IEEE computer chapter for the Waterloo-Kitchener area and editorial board for many reputed journals like Nature Scientific Reports, IET Quantum Communication, and IEEE Access. 
			\end{IEEEbiography}			
			
			
		\end{document}